\documentclass[a4paper,10pt]{revtex4-2}
\usepackage{mathrsfs}
\usepackage{graphicx}
\usepackage{latexsym}
\usepackage{amsmath}
\usepackage{amssymb}
\usepackage{textcomp}
\usepackage{amsbsy}
\usepackage{graphics}
\usepackage{epstopdf}
\usepackage{color}

\allowdisplaybreaks[4]

\begin{document}

\tolerance=5000

\title{Holographic description  of the dissipative model of universe with curvature  }

\author{I.~Brevik,$^{1}$\,\thanks{iver.h.brevik@ntnu.no}
A.~V.~Timoshkin,$^{2,3}$\,\thanks{alex.timosh@rambler.ru}
}
 \affiliation{ $^{1)}$ Department of Energy and Process Engineering,
Norwegian University of Science and Technology, N-7491 Trondheim, Norway\\
$^{2)}$Institute of Scientific Research and Development, Tomsk State Pedagogical University (TSPU),  634061 Tomsk, Russia \\
$^{3)}$ Lab. for Theor. Cosmology, International Centre  of Gravity and Cosmos,  Tomsk State University of Control Systems and Radio Electronics
(TUSUR),   634050 Tomsk, Russia
}

\tolerance=5000

\begin{abstract}
We investigate the accelerated expansion of the late-time universe in the Friedmann-Robertson-Walker metric with nonzero curvature, applying a holographic principle based on a generalized holographic dark energy model introduced by Nojiri and Odintsov  (2005,2006). We describe the evolution of the universe  using a generalized equation of state in the presence of a viscous fluid. Solutions of the gravitational equation of motions are obtained in explicit form for a constant value of the thermodynamic parameter, and for various forms of  the bulk viscosity. We calculate  analytic expressions for infrared cut-offs in terms of the particle horizon, and derive the energy conservation law in the holographic picture. We show that the inclusion of nonzero curvature in the Friedmann equation leads to the appearance of additional singularities of type Big Rip in the Universe.

\end{abstract}


\maketitle

\section{Introduction}

The discovery of the accelerated expansion of the universe \cite{1,2} led to the appearance of different cosmological models, and it turned out that cosmic acceleration can be described either via dark energy or via modified gravity \cite{3,4}. Very general dark fluid models discussed in the literature are models for which the dark energy is  described as an unconventional ideal fluid    with an inhomogeneous equation of state (EoS) \cite{5,6}.

We will be interested in universes having dark energy, in which the appearance of singularities occurs within a finite time. In general, it is necessary to consider the viscosity of the cosmic fluid  as in many situations the assumption about zero viscosity (ideal fluid) is incorrect. The influence from (bulk) viscosity affects the  behavior of the universe near cosmological singularities \cite{7,8}; cf. the Big Rip phenomena in cosmologies of type II, III and IV (what is called a Nojiri-Odintsov-Tsujikawa classification of singularities is given in \cite{9}). Also, the viscosity coefficients, even the bulk viscosity, are important in connection with turbulence effects \cite{10}.

The mathematical formalism in the present work is based upon the use of the holographic principle. It means that all information about the system parameters can be described in the form of a hologram, associated with the surface area of cosmic space \cite{11}. The generalized holographic dark energy (HDE) model, and unification of phantom inflation with  phantom acceleration, was proposed by Nojiri and Odintsov \cite{12,13}. It has later been shown that all known models of HDE fall within the Nojiri-Odintsov model \cite{14,15,16}. The development of various approaches and generalizations of HDE is given in various reviews \cite{17,18,19}. Holographic theory is shown to be in agreement with astronomical observations \cite{17,20,22,23,24,25}.

Modern measurements of the luminosity of remote objects show that the curvature of the universe is very near  zero, thus in other words  that the universe is essentially  flat. Observations from the Planck satellite  indicate that the  curvature is less than 0.03, expressed in terms of the given parameters \cite{planck2018}. This conclusion is supported by other studies also, such as in observations  of the galaxy distribution in space. Nevertheless, it is impossible to exclude by 100\% certainty that the universe has a finite curvature; cf., for instance, the discussions  in Refs.~\cite{valentino2020} and \cite{handley21}.

In this paper, we will as a working hypothesis assume that our Universe has a finite curvature, and will investigate the consequences this assumption has for the Friedmann equation. We assume the  Friedmann-Robertson-Walker (FRW) Universe. We will obtain a holographic description of cosmological models associated with an inhomogeneous viscous dark fluid, and will discuss the singular behavior of the Universe determined by this model.

\section{Main properties of the holograpic formulation}

According to the holographic principle, all physical quantities in the universe including the dark energy can be descried by specified values on the space-time boundary \cite{26,27}. The typical HDE density can be described via the Planck mass $M_p$ and a characteristic length $L_{IR}$ (infrared radius) \cite{11},
\begin{equation}
\rho_{\rm hol}=3c^2 M_p^2
L_{\rm IR}^{-2}, \label{1}
\end{equation}
where $c$ is an arbitrary dimensionless positive parameter (we consider an expanding universe).

Let us consider the  homogeneous and isotropic FRW universe with metric
\begin{equation}
ds^2= -dt^2+a^2(t)\left( \frac{dr^2}{1-\lambda r^2} +r^2d\Omega^2\right),  \label{2}
\end{equation}
where $a(t)$ is the scale factor and $d\Omega^2=d\theta^2+\sin^2 \theta d\varphi^2$ is the metric of the space. The quantity $\lambda$ characterizes the curvature of the three-dimensional space.

As is known, the metric (\ref{2}) describes a homogeneous and isotropic expanding space. We may have $\lambda=0$ (spatially flat space), $\lambda=+1$ (closed universe), or $\lambda =-1$ (open universe). The open universe expands forever; the flat universe also expands forever but at $t\rightarrow +\infty$ the expansion occurs at constant speed; the closed universe expands to a certain instant, after which the expansion is replaced by a compression leading to a collapse.

The  Friedmann equation for a one-component fluid with nonzero curvature has the form
\begin{equation}
H^2 = \frac{k^2}{3}\rho -\frac{\lambda}{3},  \label{3}
\end{equation}
where $\rho$ is the HDE density, $k^2=8\pi G$ with $G$ the Newtonian gravitational constant, and
 $H(t)= \dot{a}(t)/a(t)$ is the Hubble function.

Currently there are no strong recommendations for how the infrared radius $L_{IR}$ should be chosen. One may identify this parameter either with the particle horizon $L_p$ or with the event horizon $L_f$ \cite{12},
\begin{equation}
L_p(t)= a(t)\int_0^t \frac{dt'}{a(t')}, \quad L_f(t)= a(t)\int_t^\infty \frac{dt'}{a(t')}. \label{4}
\end{equation}
In general, the infrared cut-off $L_{IR}$ could be a combination of the parameters $L_p,L_f$ and their derivatives. It could also contain the Hubble function, the scale factor, and its derivatives \cite{12}.

It should be noted, however,  that not all ways of choosing infrared radius will lead to an accelerated
 expansion of the universe so that the choice of an infrared radius is not arbitrary.

 If we identify the energy density $\rho$ in Eq.~(\ref{3}) with the HDE quantity $\rho_{\rm hol}$ in Eq.~(\ref{1}), we obtain the first    Friedmann equation in another form,
 \begin{equation}
 H= \sqrt{ \left( \frac{c}{L_{IR}}\right)^2-\frac{\lambda}{a^2}}. \label{5}
 \end{equation}
In the following we will assume that the viscous dark fluid, driving the evolution of the universe, has a holographic origin.

\section{Holography of viscous fluid models in space with nonzero curvature}

In this section we will consider cosmological models of the viscous dark fluid, where the fluid satisfies the inhomogeneous equation of state in a FRW universe \cite{5,29},
Let us consider a viscous dark fluid with an effective inhomogeneous equation of state (EoS) in flat FRW space-time,
\begin{equation}
p= \omega (\rho,t)\rho -3H\zeta(H,t), \label{6}
\end{equation}
where $\omega(\rho,t)$ is the thermodynamic parameter,   $\zeta(H,t)$ is the bulk viscosity. For thermodynamic reasons, $\zeta(H,t)>0$. Dissipative processes are described by the bulk viscosity. We assume the following form for it \cite{30},

\begin{equation}
 \zeta(H,t)= \xi_1(t)(3H)^n, \label{7}
 \end{equation}
 with $n>0$.

 Let us assume that the universe is filled with a one-component fluid, obeying the energy conservation law
\begin{equation}
\dot{\rho}+3H(\rho+p)=0. \label{8}
\end{equation}
We apply the holographic principle for cosmological models with a constant value of the thermodynamic parameter $\omega (\rho,t))=\omega_0$ and various values of the bulk viscosity $\zeta(H,t)$, and distinguish between two cases.

\bigskip

\noindent {\it Case 1. Fluid model with constant $\omega(\rho,t)=\omega_0$ and constant viscosity $\zeta(H,t)=\zeta_0$.}

\bigskip
\noindent The equation of state (\ref{6}) takes the form
\begin{equation}
p=\omega_0\rho -3\zeta_0 H. \label{9}
\end{equation}
Next, using equations (\ref{3}), (\ref{8}) and (\ref{9}) one obtains the following differential equation relating the scale factor to the energy density,
\begin{equation}
a\ddot{a}+(\tilde{\omega}_0-1)\dot{a}^2-\tilde{\zeta}_0a\dot{a}+\lambda (\tilde{\omega}_0+1)=0, \label{10}
\end{equation}
where $\tilde{\omega}_0=\frac{3}{2}(\omega_0+1)$ and $\tilde{\zeta}_0=\frac{3}{2}k^2\zeta_0$.
The solution of this equation is
\begin{equation}
a(t)= \sqrt{C_1+C_2e^{\tilde{\zeta}_0t} +\theta t}, \label{11}
\end{equation}
where $\theta = \frac{\lambda}{\tilde{\zeta}_0}(\tilde{\omega}_0+1)$ and $C_1,C_2$ are arbitrary constants. The expression (\ref{11}) for the scale factor contains the correction $\theta t$, which is associated with the spatial curvature.

 For the Hubble function we obtain the expression
\begin{equation}
H(t)= \frac{1}{2}\frac{ C_2\tilde{\zeta}_0 e^{\tilde{\zeta}_0t}+\theta}{C_1+C_2e^{ \tilde{\zeta}_0t} +\theta t}. \label{12}
\end{equation}
If $C_1$ and $C_2$ are positive, then in a flat universe $(\theta=0)$ there is no singularity. In an open universe $(\theta <0)$ it is possible to form a singularity after a finite time span (type Big Rip \cite{6}).

Let us consider the asymptotic behavior of the universe. In the limit $t\rightarrow 0$ (inflationary epoch) the Hubble function tends to the constant value $H(t)= \frac{C_2\tilde{\zeta}_0+\theta}{C_2\tilde{\zeta}_0 +C_1}$, while in the limit $t\rightarrow \infty$ (dark energy epoch) it goes again to a constant $H(t)\rightarrow \tilde{\zeta}_0$ regardless of the spatial curvature. In both cases, the universe is in a state of accelerated expansion.

Now calculate the particle horizon $L_p$ for nonzero spatial curvature,
\begin{equation}
L_p= \frac{1}{\theta}\int_{C_1}^{\theta t+C_1} \frac{dx}{\sqrt{\alpha e^{\beta x}+x}}, \label{13}
\end{equation}
where $\alpha= C_2e^{-C_1\beta}, \, \beta= \frac{\tilde{\zeta}_0}{\theta}$ and $\theta \neq 0$.

In the particular case when $C_1=0, C_2=1 ~(\alpha=1)$, we obtain in the initial stage inflation
\begin{equation}
L_p(t\rightarrow 0)=\frac{2}{\gamma}\sqrt{1+\gamma t}\,(\sqrt{1+\gamma t}-1), \label{14}
\end{equation}
where $\gamma = \theta + \tilde{\zeta}_0$.

The Hubble parameter $H$ can be expressed in terms of the particle horizon as
\begin{equation}
H= \frac{{\dot{L}}_p-1}{L_p}, \quad \dot{H}= \frac{{\ddot{L}}_p}{ L_p}- \frac{{\dot{L}}_p^2}{L_p^2} + \frac{{\dot{L}}_p}{L_p^2}. \label{15}
\end{equation}
Then,  by using (\ref{15}) the energy conservation equation (\ref{8}) can be rewritten as
\begin{equation}
 \frac{{\ddot{L}}_p}{ L_p}- \frac{{\dot{L}}_p^2}{L_p^2} + \frac{{\dot{L}}_p}{L_p^2}+\tilde{\omega}_0
 \left( \frac{\dot{L}_p-1}{L_p}\right)^2
 +\left[2\lambda (\tilde{\omega}_0+1)(C_2\tilde{\zeta}_0e^{\tilde{\zeta}_0t}+\theta)-\tilde{\zeta}_0\right]
  \frac{{\dot{L}}_p -1}{L_p} =0. \label{16}
 \end{equation}
 In flat space $(\lambda=0)$ this equation simplifies to
 \begin{equation}
  \frac{{\ddot{L}}_p}{ L_p}- \frac{{\dot{L}}_p^2}{L_p^2} + \frac{{\dot{L}}_p}{L_p^2}+\tilde{\omega}_0
 \left( \frac{\dot{L}_p-1}{L_p}\right)^2 -\tilde{\zeta}_0
  \frac{{\dot{L}}_p -1}{L_p} =0. \label{17}
 \end{equation}
 Thus, we have obtained an energy conservation equation, based on a dissipative model of the viscous holographic fluid.

 \bigskip

\noindent {\it Case 2. Fluid model with constant $\omega(\rho,t)=\omega_0$ and viscosity proportional to the Hubble function, $\zeta(H,t)=3\tau H$.}

\bigskip

Assume that $\omega(\rho,t)=\omega_0$ as before, and take now the bulk viscosity to be a linear function of the Hubble parameter, $\zeta(H,t)=3\tau H$, where $\tau$ is a positive dimensional constant. Then the equation of state (\ref{6}) takes the form
\begin{equation}
p= \omega_0\rho -9\tau H^2. \label{18}
\end{equation}
Using equations (\ref{3}), (\ref{8}) and (\ref{17}), we can represent the energy conservation law as a differential equation for the scale factor,
\begin{equation}
a \ddot{a}+(\tilde{\tau}-1)\dot{a}^2+ \tilde{\lambda}=0, \label{19}
\end{equation}
where $\tilde{\tau}= \frac{3}{2}(\omega_0+1-3\tau k^2)$ is a modified viscosity parameter and $\tilde{\lambda}=\frac{1}{2}\lambda(3\tilde{\omega_0}+5)$ a modified curvature parameter.

In this case the expression for the scale factor becomes
\begin{equation}
a(t)= \sqrt{ C_1t+C_2-\frac{1}{2}\tilde{\lambda}\tilde{\tau}t^2}, \label{20}
\end{equation}
where $C_1,C_2$ are arbitrary constants.

The Hubble function is
\begin{equation}
H(t)= \frac{1}{2}\frac{
C_1-\tilde{\lambda}\tilde{\tau}t}{C_1t+C_2-\frac{1}{2}\tilde{\lambda}\tilde{\tau}t^2}. \label{21}
\end{equation}
If $C_1 \neq 0$ and $C_2\neq 0$ in a space with nonzero curvature, there will be two singularities of the Big Rip type, while in the case of flat space, only one singularity is formed.Thus, the curvature of space leads to additional singularities.

In the asymptotic cases of early or late universe the Hubble function approaches a constant, independent of the curvature.

In order to deal with the holographic description, we identify the infrared radius with the particle horizon \cite{12}. A calculation of (\ref{4}) in the case when $C_1=0, C_2=1$ leads to the result
\begin{equation}
L_p= \sqrt{\frac{2}{\tilde{\lambda}\tilde{\tau}}} \sqrt{1-\frac{1}{2}\tilde{\lambda}\tilde{\tau}t^2}\, \arcsin \left(  \sqrt{\frac{1}{2}\tilde{\lambda}\tilde{\tau}t}\right). \label{22}
\end{equation}
For the spatially flat universe the size of the particle horizon becomes
\begin{equation}
L_p= \frac{2}{C_1}\sqrt{C_1t+C_2}\, (\sqrt{C_1t+C_2} -\sqrt{C_2}). \label{23}
\end{equation}
In this case, the energy conservation law (\ref{8}) in holographic form is
\begin{equation}
  \frac{{\ddot{L}}_p}{ L_p}- \frac{{\dot{L}}_p^2}{L_p^2} + \frac{{\dot{L}}_p}{L_p^2}+\tilde{\tau}
 \left( \frac{\dot{L}_p-1}{L_p}\right)^2 +2\tilde{\lambda}(C_1-\tilde{\tau}\tilde{\lambda}t)
  \frac{{\dot{L}}_p -1}{L_p} =0. \label{24}
 \end{equation}
 Concluding this section, we have  demonstrated the possibility of using the holographic principle to describe the evolution of the universe with a nonzero curvature within the frames of standard viscous cosmology.

 \section{Conclusion}

 In this article we have studied the dissipative model of the expanding universe under the assumption of a homogeneous and isotropic Friedmann-Robertson-Walker metric with nonzero spatial curvature. To investigate the universe's evolution a holographic principle was adopted, on the basis of the generalized holographic dark energy model proposed by Nojiri and Odintsov \cite{12,13}. Within the framework of this model, by using the generalized equation of state for the dark viscous fluid, the influence from the spatial curvature on the singular behavior of the universe was carried out. The holograhic energy density is known to be basically the same as the energy of the infrared (IR) radiation. The infrared radius $L_{IR}$ can be chosen in various ways, typically either as the particle horizon or the event horizon;  cf. Eq.~(\ref{4}). In Sec.~III we investigated two different models for the bulk viscosity. For both models, analytic expressions were obtained for the energy conservation law in terms of the particle horizon $L_p$.

 Although experimental evidence show that the observed universe is spatially flat to a good accuracy, it is of interest to notice that for a  nonzero curvature one obtains extra singularities in addition to those associated with  a flat universe.

 It ought also to be mentioned that applications of the holographic method have led to theoretical predictions in good agreement  with astronomical observations \cite{31}.

\section*{Acknowledgment}

This work was supported by Russian Fund for Fundamental Studies; Project No. 20-52-05009  (A. V. T.).

\end{document}